\documentclass{amsart}

\usepackage{enumerate, amsmath, amsfonts, amssymb, amsthm, wasysym, graphics, graphicx, xcolor, url, hyperref, hypcap, multirow}
\hypersetup{colorlinks=true, citecolor=darkblue, linkcolor=darkblue}
\graphicspath{{figures/}}

\title[Enumerating topological $(n_k)$-configurations]{Enumerating topological $(n_k)$-configurations}

\thanks{Vincent Pilaud was partially supported by grant MTM2011-22792 of the Spanish Ministerio de Ciencia e Innovaci\'on and by European Research Project ExploreMaps (ERC StG 208471).}

\author{J\"urgen Bokowski}
\address{Technische Universit\"at Darmstadt}
\email{juergen.bokowski@googlemail.com}

\author{Vincent Pilaud}
\address{CNRS \& LIX, \'Ecole Polytechnique, Palaiseau}
\email{vincent.pilaud@lix.polytechnique.fr}
\urladdr{http://www.lix.polytechnique.fr/~pilaud/}

\newtheorem{resultat}{Result}[]
\newtheorem{probleme}[resultat]{Problem}

\newcommand{\N}{\mathbb{N}}
\newcommand{\Z}{\mathbb{Z}}
\newcommand{\Q}{\mathbb{Q}}

\newcommand{\bP}{\mathbb{P}}
\newcommand{\fS}{\mathfrak{S}}
\newcommand{\cC}{\mathcal{C}}
\newcommand{\partitions}{\Pi}
\newcommand{\bb}[1]{\mathbf{\darkblue #1}}

\newcommand{\multiset}[2]{\left\{\!\!\left\{ #1 \;\middle|\; #2 \right\}\!\!\right\}}

\newcommand{\ie}{\textit{i.e.}~} 
\newcommand{\eg}{\textit{e.g.}~} 

\newcommand{\conf}[2]{\mbox{$(#1_#2)$-confi}\-gu\-ra\-tion}
\newcommand{\cross}[1]{\mbox{$#1$-crossing}}

\newcommand{\Maple}{\textsc{maple}}
\newcommand{\Java}{\textsc{java}}
\newcommand{\Haskell}{\textsc{haskell}}
\newcommand{\Cinderella}{\textsc{cinderella}}

\definecolor{darkblue}{rgb}{0,0,0.7}
\newcommand{\darkblue}{\color{darkblue}}
\newcommand{\defn}[1]{\emph{\darkblue #1}}
\graphicspath{{figures/}}
\renewcommand{\paragraph}[1]{\vspace{.3cm}\noindent{\sc #1} --- }

\begin{document}

\begin{abstract}
An \conf{n}{k} is a set of $n$ points and $n$~lines in the projective plane such that their point\,--\,line incidence graph is $k$-regular. The configuration is geometric, topological, or combinatorial depending on whether lines are considered to be straight lines, pseudolines, or just combinatorial lines.

We provide an algorithm for generating, for given~$n$ and~$k$, all topological \conf{n}{k}s up to combinatorial isomorphism, without enumerating first all combinatorial \conf{n}{k}s. We apply this algorithm to confirm efficiently a former result on topological \conf{18}{4}s, from which we obtain a new geometric \conf{18}{4}. Preliminary results on \conf{19}{4}s are also briefly reported.
\end{abstract}

\maketitle


\section{Introduction}

A \defn{point\,--\,line configuration} is a set $P$ of \defn{points} and a set $L$ of \defn{lines} together with an \defn{incidence relation}, where two points of $P$ can be incident with at most one line of~$L$ and two lines of~$L$ can be incident with at most one point of $P$. Throughout the paper, we only consider \defn{connected} configurations, where any two elements of~$P \sqcup L$ are connected via a path of incident elements. An \defn{isomorphism} (resp.~a \defn{duality}) between two configurations $(P,L)$ and $(P',L')$ is an incidence-preserving map from~$P \sqcup L$ to $P' \sqcup L'$ which sends points to points and lines to lines (resp.~ which exchanges points and lines).

According to~the underlying structure, we distinguish three different levels of configurations, in increasing generality:

\begin{description}
\item[\rm\defn{Geometric configuration}] Points and lines are points and lines in the real projective plane~$\bP$.
\item[\rm\defn{Topological configuration}] Points are points in~$\bP$, but lines are \defn{pseudolines}, \ie non-separating simple closed curves of~$\bP$.
\item[\rm\defn{Combinatorial configuration}] Just an abstract incidence structure $(P,L)$ as described above, with no additional geometric structure.
\end{description}

\begin{figure}
  \centerline{\includegraphics[width=1\textwidth]{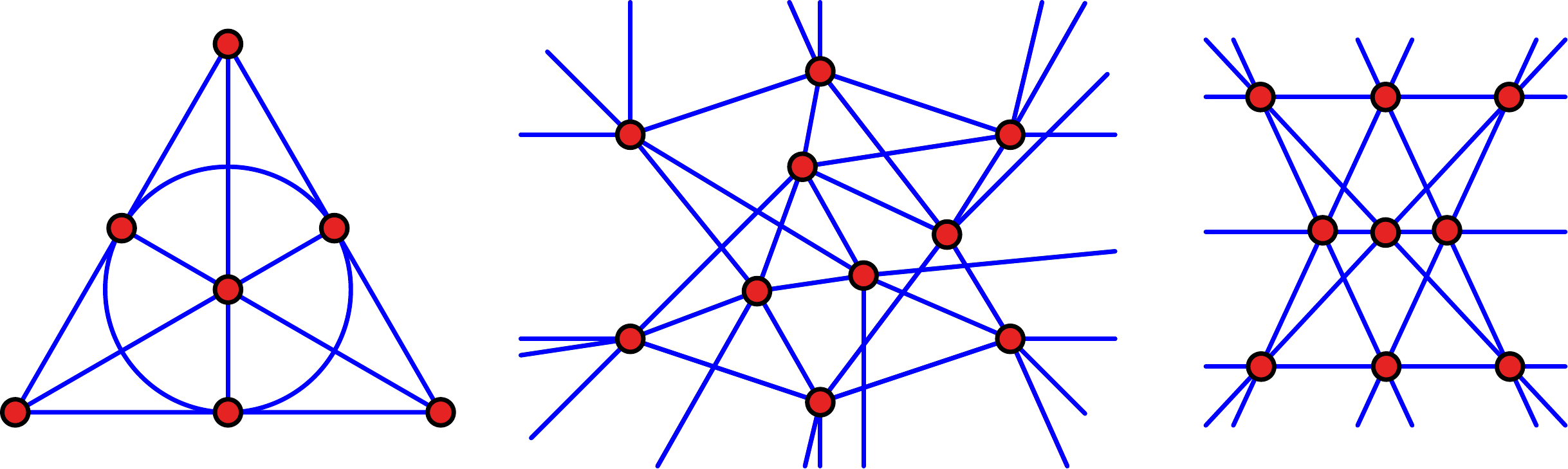}}
  \caption{(Left) Fano's configuration is a combinatorial \conf{7}{3} but is not realizable topologically or geometrically. (Center) Kantor's topological \conf{10}{3} is not realizable geometrically. (Right) Pappus' configuration is a geometric \conf{9}{3}.}
  \label{fig:famousConfigurations}
\end{figure}

In this paper, we focus on \defn{regular} configurations, \ie whose incidence relation is regular. More precisely, an \defn{\conf{n}{k}}~$(P,L)$ is a set $P$ of $n$ points and a set~$L$ of $n$ lines such that each point of $P$ is contained in $k$ lines of $L$ and each line of $L$ contains $k$ points of $P$.
We have represented three famous $3$-regular configurations in Figure~\ref{fig:famousConfigurations} to illustrate the previous definitions.

Point\,--\,line configurations have a long history in discrete $2$-dimensional geometry. We refer to Branko Gr\"unbaum's  recent monograph~\cite{Grunbaum1} for a detailed treatment of the topic and for historical references. As underlined in this monograph, the current study of regular configurations focusses on the following two problems:
\begin{enumerate}[(i)]
\item For a given~$k$, determine for which values of~$n$ do geometric, topological, and combinatorial \conf{n}{k}s exist.
\item Enumerate and classify \conf{n}{k}s for given $k$ and $n$.
\end{enumerate}
In particular, it is challenging to determine the minimal value $n$ for which \conf{n}{k}s exist and to enumerate these minimal configurations.

For $k \in \{3,4\}$, the existence of \conf{n}{k}s is almost completely understood.
When $k=3$, combinatorial \conf{n}{3}s exist for every $n\ge7$, but topological and geometric \conf{n}{3}s exist only for every $n\ge9$.
When $k=4$, combinatorial \conf{n}{4}s exist iff $n\ge13$, topological \conf{n}{4}s exist iff $n\ge17$~\cite{BokowskiSchewe1, BokowskiGrunbaumSchewe} and geometric \conf{n}{4}s exist iff $n\ge18$~\cite{Grunbaum4, BokowskiSchewe2}, with the possible exceptions of $19$, $22$, $23$, $26$, $37$ and~$43$. For~$k \ge 5$, the situation is more involved, and the existence of combinatorial, topological and geometric \conf{n}{k}s is not determined in general.

Concerning the enumeration, an important effort has been done on combinatorial $(n_3)$- and \conf{n}{4}s. Table~\ref{table:comb} provides the known values of the number $c_k(n)$ of combinatorial \conf{n}{k}s up to isomorphism.
The first row of this table ($k=3$) appeared in~\cite{BettenBrinkmannPisanski}, except $c_3(19)$ which was announced later on in~\cite[p.275]{PBMOG}.
The second row ($k=4$) appeared in~\cite[p.34]{BettenBetten}, except $c_4(19)$ which was only computed recently in~\cite{PaezOsunaSanAugustinChi}.

\begin{table}[b]
\centerline{
$
\begin{array}{c|cccccccccccccc}
n & \le 6 & 7 & 8 & 9 & 10 & 11 & 12 & 13 & 14 & 15 & 16 & 17 & 18 & 19 \\
\hline
c_3(n) & 0 & 1 & 1 & 3 & 10 & 31 & 229 & 2\,036 & 21\,399 & 245\,342 & 3\,004\,881 & 38\,904\,499 & 530\,452\,205 & 7\,640\,941\,062 \\
c_4(n) & 0 & 0 & 0 & 0 & 0 & 0 & 0 & 1 & 1 & 4 & 19 & 1\,972 & 971\,171 & 269\,224\,652
\end{array}
$}

\medskip
\caption{The number $c_k(n)$ of combinatorial \conf{n}{k}s up to isomorphism.}
\label{table:comb}
\end{table}

In this paper, we are interested in the numbers $t_k(n)$ and $g_k(n)$ of topological and geometric \conf{n}{k}s up to isomorphism. To obtain these numbers, one method is to select the topologically or geometrically realizable configurations among the list of all combinatorial \conf{n}{k}s. For example, the numbers $t_3(n)$ and $g_3(n)$ presented in Table~\ref{table:topo&geom} were derived from a careful study of the corresponding combinatorial configurations (see the historical remarks and references in~\cite{Grunbaum1}).
In~\cite{Schewe}, Lars Schewe provided a general method to study the topological realizability of a combinatorial configuration using satisfiability solvers, and obtained the numbers $t_4(17)=1$ and $t_4(18)=16$.
In~\cite{BokowskiSchewe2}, J\"urgen Bokowski and Lars Schewe studied the geometric realizability of a combinatorial configuration. This question is clearly an instance of the existential theory of the reals (ETR): it boils down to determining whether a set of polynomial equalities and inequalities admits a solution in the reals (indeed, the inclusion of a point in a line can be tested by a polynomial equation). Using the construction sequences presented in~\cite{BokowskiSchewe2}, the complexity of this instance of ETR can be decreased significantly. With this method, J\"urgen Bokowski and Lars Schewe showed that the only combinatorial \conf{17}{4} which is topologically realizable is not geometrically realizable and they exhibited a geometric \conf{18}{4}.

Table~\ref{table:topo&geom} summarizes the values of $t_3(n), g_3(n), t_4(n)$ and $g_4(n)$ known up-to-date (we have additionally included our results in bold letters; see below). This table indicates a clear difference of behavior between $3$- and $4$-regular configurations. On the one hand, when $k=3$, most of the combinatorial \conf{n}{3}s are topologically and geometrically realizable for small values of~$n$. For~$n \le 12$, the only counter-examples are the Fano \conf{7}{3}, the M\"obius-Kantor \conf{8}{3}, and Kantor's \conf{10}{3} ---~see Figure~\ref{fig:famousConfigurations} (left \& center). On the other hand, when $k=4$, it is not reasonable to look for geometric \conf{n}{4}s among all combinatorial \conf{n}{4}s. To further extend our knowledge on geometric configurations, it thus seems crucial to limit our research to those combinatorial configurations which are already topologically realizable.

\begin{table}
$$
\begin{array}{c|ccc}
n & c_3(n) & t_3(n) & g_3(n) \\
\hline
\le 6 & 0 & 0 & 0 \\
7 & 1 & 0 & 0 \\
8 & 1 & 0 & 0 \\
9 & 3 & 3 & 3 \\
10 & 10 & 10 & 9 \\
11 & 31 & 31 & 31 \\
12 & 229 & 229 & 229 \\
13 & 2\,036 & ? & ?
\end{array}
\qquad\qquad
\begin{array}{c|ccc}
n & c_4(n) & t_4(n) & g_4(n) \\
\hline
\le 12 & 0 & 0 & 0 \\
13 & 1 & 0 & 0 \\
14 & 1 & 0 & 0 \\
15 & 4 & 0 & 0 \\
16 & 19 & 0 & 0 \\
17 & 1\,972 & 1 & 0 \\
18 & 971\,191 & 16 & \bb{2} \\
19 & 269\,224\,652 & \bb{4\,028} & ?
\end{array}
$$
\caption{The numbers $t_k(n)$ of topological \conf{n}{k}s and $g_k(n)$ of geometric \conf{n}{k}s up to isomorphism.}
\label{table:topo&geom}
\end{table}

Motivated by this observation, we present an algorithm for generating, for given $n$ and $k$, all topological \conf{n}{k}s up to isomorphism, without enumerating first all combinatorial \conf{n}{k}s. The algorithm sweeps the projective plane to construct a topological \conf{n}{k}~$(P,L)$, but only considers as relevant the events corresponding to the sweep of points of~$P$. This strategy enables us to identify along the way some isomorphic topological configurations, and thus to maintain a reasonable computation space and time.

We have developed two different implementations of this algorithm. The first one was written in \Haskell{} by the first author to develop the strategy of the enumeration process. Once the general idea of the algorithm was settled, the second author wrote another implementation in \Java{}, focusing on the optimization of computation space and time of the process.

We outline three applications of our algorithm. 
First, the algorithm is interesting in its own right. Before describing some special methods for constructing topological configurations, Branko~Gr\"unbaum writes in \mbox{\cite[p.\,165]{Grunbaum1}} that \emph{``the examples of topological configurations presented so far have been ad hoc, obtained essentially through (lots of) trial and error''}. Our algorithm can reduce considerably the \emph{``trial and error''} method.
Second, our algorithm enables us to check and confirm all values of $t_4(n)$, for $n\le 18$, obtained in earlier papers. We can use for that a single method and reduce considerably the computation time (\eg the computation of the \conf{18}{4}s needed several months of CPU-time in~\cite{Schewe}, and only one hour with our \Java{} implementation).
Finally, this algorithm enables us to compute all $t_4(19) = 4028$ isomorphism classes of topological \conf{19}{4}s.

As an application of our enumeration results, we studied in detail the possible geometric realizations of the topological \conf{18}{4}s. Using a \Maple{} inplementation of the construction sequence method of J\"urgen Bokowski and Lars Schewe~\cite{BokowskiSchewe1}, we obtain that there are precisely $2$ geometric \conf{18}{4}s: the first \conf{18}{4} constructed in~\cite{BokowskiSchewe1}, plus an additional one which appears for the first time in this paper. In contrast, deriving the list of geometric \conf{19}{4}s from the list of topological \conf{19}{4}s still requires some computational effort and is left to a subsequent paper.

The first section of this paper is devoted to the enumeration algorithm for isomorphism classes of topological configurations. The second section presents the application to the enumeration of geometric \conf{18}{4}s.

Topological configurations are pseudoline arrangements, or rank~$3$ oriented matroids. We assume the reader to have some basic knowledge on these topics. We refer to~\cite{Bokowski,BVSWZ,Knuth} for introductions.


\section{Topological configurations}
\label{sec:topoconf}

In this section, we present our algorithm to generate all isomorphism classes of topological \conf{n}{k}s, for given $n$ and $k$. Let us insist again on the crucial fact that we do not need to enumerate first all combinatorial \conf{n}{k}s. The main idea of the algorithm is to sweep the projective plane to construct a topological \conf{n}{k}~$(P,L)$, only focussing on the ``relative positions of the points of $P$'' and ignoring at first the ``relative positions of the other crossings of the pseudolines of $L$'' (precise definitions are given in Section~\ref{subsec:equivrelations}). This strategy enables us to identify along the way some isomorphic topological configurations, and thus to maintain a reasonable computation space and time.

\subsection{Three equivalence relations}
\label{subsec:equivrelations}

There are three distinct notions of equivalence on topological configurations.

The finest notion is the usual notion of topological equivalence between pseudoline arrangements in the projective plane: two configurations are \defn{topologically equivalent} if there is an homeomorphism of their underlying projective planes that sends one arrangement onto the other. 

The coarsest notion is that of combinatorial equivalence: two \conf{n}{k}s are \defn{combinatorially equivalent} if they are isomorphic as combinatorial \conf{n}{k}s.

The intermediate notion is based on the graph of admissible mutations. Remember that a \defn{mutation} in a pseudoline arrangement is a local transformation of the arrangement where only one pseudoline $\ell$ moves, sweeping a single vertex $v$ of the remaining arrangement. It only changes the position of the crossings of~$\ell$ with the pseudolines incident to $v$. If those crossings are all \cross{2}s, the mutation does not perturb the \cross{k}s of the arrangement, and thus produces another topological \conf{n}{k}. We say that such a mutation is \defn{admissible}. Two configurations are \defn{mutation equivalent} if one can be obtained from the other by a (possibly empty) sequence of admissible mutations followed by an homeomorphism of the underlying projective space.

\begin{figure}[h]
  \centerline{\includegraphics[scale=.75]{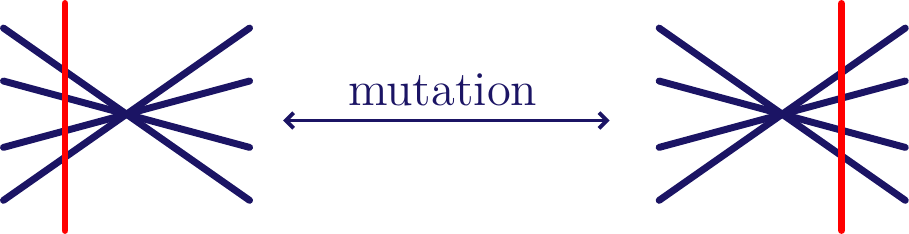}}
  \caption{An admissible mutation.}
  \label{fig:mutation}
\end{figure}

Obviously, topological equivalence implies mutation equivalence, which in turn implies combinatorial equivalence. The reciprocal implications are wrong.

\begin{figure}[h]
  \centerline{\includegraphics[width=1.3\textwidth]{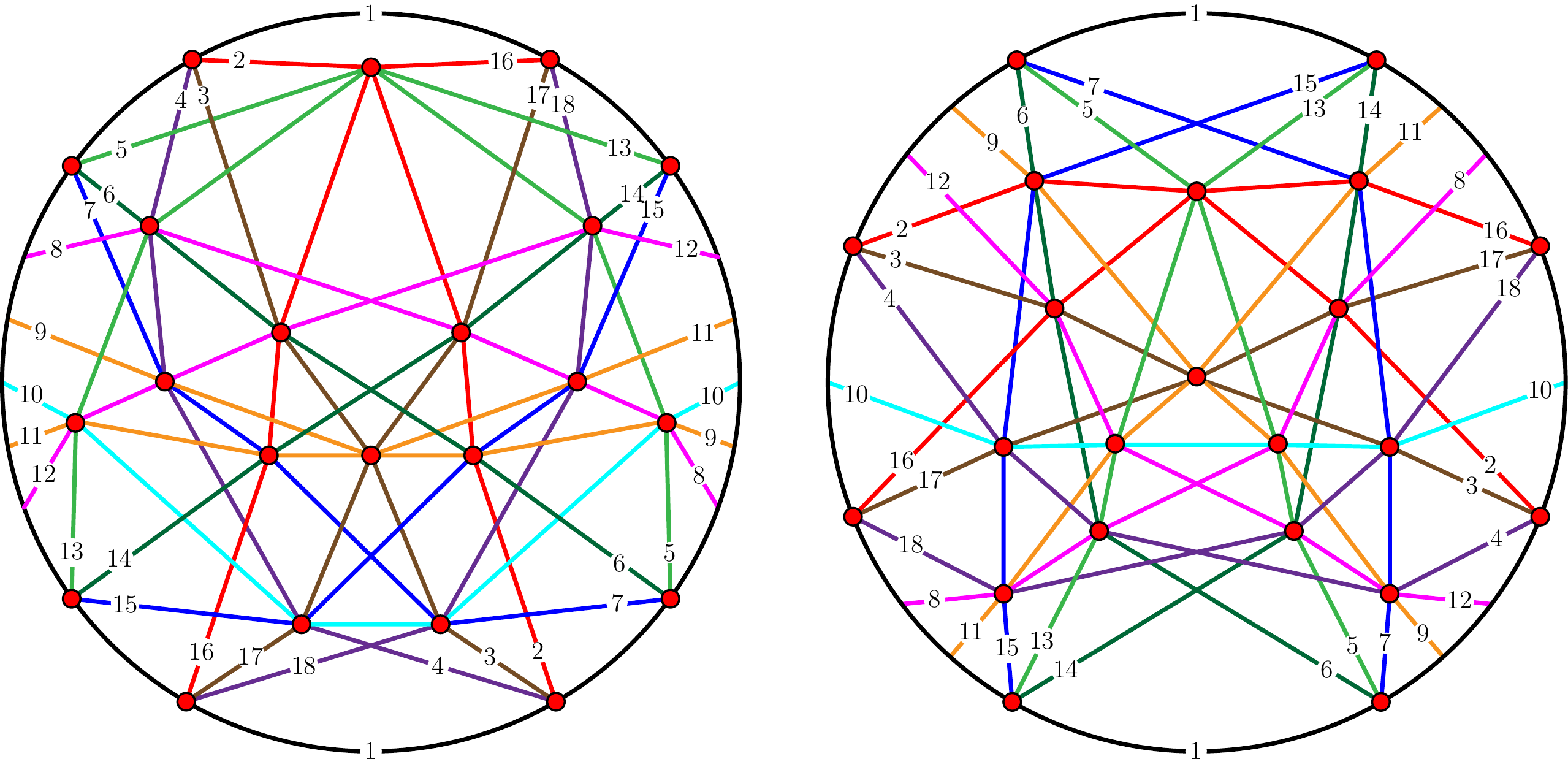}}
  \caption{Two \conf{18}{4}s which are combinatorially equivalent but neither mutation nor topologically equivalent.}
  \label{fig:configurations_18_4}
\end{figure}

As an illustration, the two \conf{18}{4}s depicted in Figure~\ref{fig:configurations_18_4} are combinatorially equivalent (the labels on the pseudolines provide a combinatorial isomorphism) but not topologically equivalent (the left one has $22$ quadrangles and $2$ pentagons, while the right one has $23$ quadrangles). In fact, one can even check that they are not mutation equivalent.

\subsection{Representation of arrangements}
\label{subsec:representation}

In this section, we state certain properties of configurations that we can assume without loss of generality. In particular, we choose a suitable representation of our pseudoline arrangements that we will use for the description of the algorithm in Section~\ref{subsec:description}.

\paragraph{Simple configurations}
A topological configuration $(P,L)$ is \defn{simple} if no three pseudolines of $L$ meet at a common point except if it is a point of~$P$. Since any topological \conf{n}{k} can be arbitrarily perturbed to become simple, we only consider simple topological \conf{n}{k}s. Once we obtain all simple  topological \conf{n}{k}s, it is usual to obtain all (non-necessarily simple) topological \conf{n}{k}s up to topological equivalence by exploring the mutation graph, and we do not report on this aspect.

In a simple \conf{n}{k} $(P,L)$, there are two kinds of intersection points among pseudolines of $L$: the points of $P$, which we also call \defn{\cross{k}s}, and the other intersection points, which we call \defn{\cross{2}s}. Each pseudoline of~$L$ contains $k$ \cross{k}s and $n-1-k(k-1)$ \cross{2}s. In total, a simple \conf{n}{k} has $n$ \cross{k}s and ${n \choose 2} - n({k \choose 2}-1)$ \cross{2}s.

\paragraph{Segment length distributions}
A \defn{segment} of a topological configuration~$(P,L)$ is the portion of a pseudoline of~$L$ located between two consecutive points of~$P$. If~$(P,L)$ is simple, a segment contains no \cross{k} except its endpoints, but may contain some \cross{2}s. The \defn{length} of a segment is the number of \cross{2}s it contains.

The circular sequence of the segment lengths on a pseudoline of $L$ forms a \mbox{$k$-partition} of ${n-1-k(k-1)}$. 
We call a \defn{maximal representative} of a $k$-tuple the lexicographic maximum of its orbit under the action of the dihedral group (\ie rotations and reflections of the $k$-tuple). We denote by $\partitions$ the list of all distinct maximal representatives of the $k$-partitions of $n-1-k(k-1)$, ordered lexicograhically. For example, when $k=4$ and $n=17$, we have $\partitions = [4,0,0,0]$, $[3,1,0,0]$, $[3,0,1,0]$, $[2,2,0,0]$, $[2,0,2,0]$, $[2,1,1,0]$, $[2,1,0,1]$, $[1,1,1,1]$.

\begin{figure}
  \centerline{\includegraphics[width=1.45\textwidth]{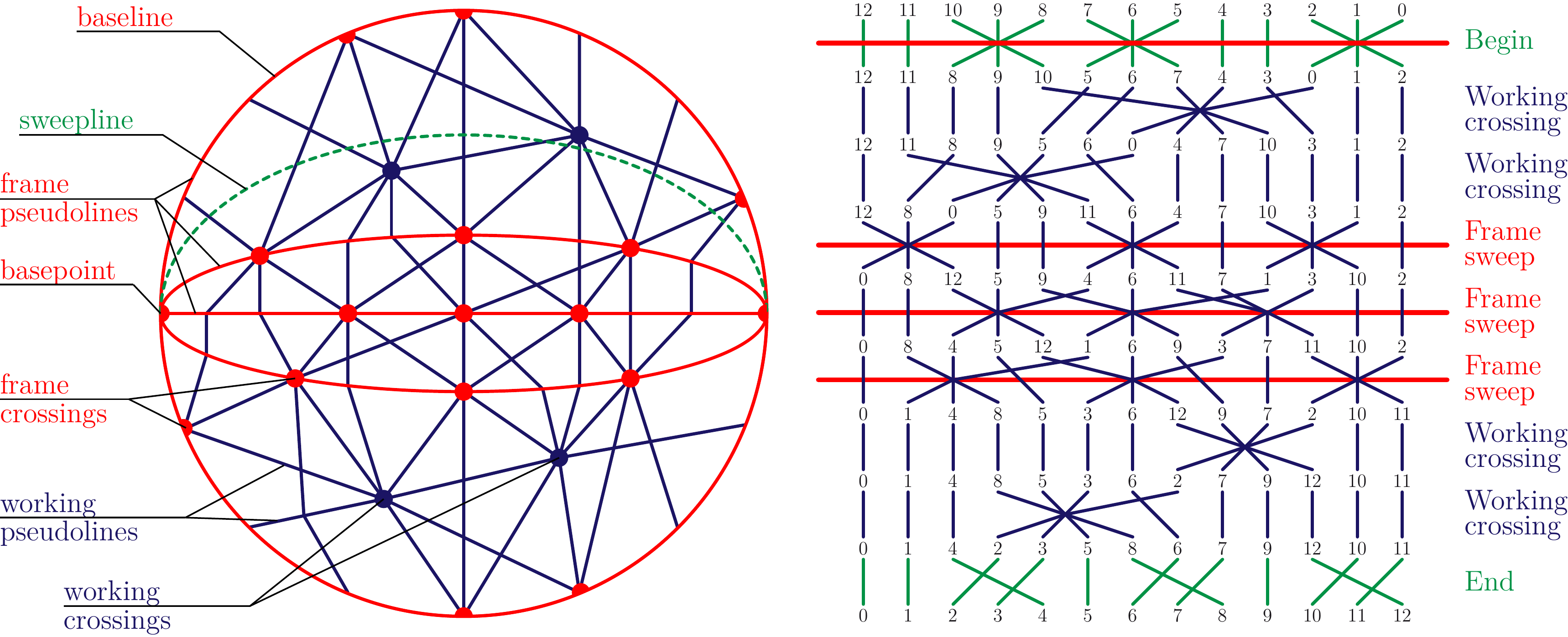}}
  \caption{Suitable representation of a \conf{17}{4}, and the corresponding wiring diagram.}
  \label{fig:configuration}
\end{figure}

\paragraph{A suitable representation}
We represent the projective plane as a disk where we identify antipodal boundary points. Given a simple topological \conf{n}{k}~$(P,L)$, we fix a representation of its underlying projective plane which satisfies the following properties (see Figure~\ref{fig:configuration} left).

The leftmost point of the disk (which is identified with the rightmost point of the disk) is a point of~$P$, which we call the \defn{base point}. 
The $k$ pseudolines of~$L$ passing through the base point are called the \defn{frame pseudolines}, while the other $n-k$ pseudolines of $L$ are called \defn{working pseudolines}.
The frame pseudolines decompose the projective plane into $k$ connected regions which we call \defn{frame regions}.
A crossing is a \defn{frame} crossing if it involves a frame pseudoline and a \defn{working} crossing if it involves only working pseudolines.

The boundary of the disk is a frame pseudoline, which we call the \defn{base line}. We furthermore assume that the segment length distribution $\Lambda$ on the top half-circle appears in $\partitions$ (\ie is its own maximal representative), and that no maximal representative of the segment length distribution of a pseudoline of $L$ appears before $\Lambda$ in $\partitions$. In particular, the leftmost segment of the base line is a longest segment of the configuration.

\paragraph{Wiring diagram and allowable sequence}
Another interesting representation of our \conf{n}{k} is the \defn{wiring diagram}~\cite{GoodmanPollack} of its working pseudolines (see Figure~\ref{fig:configuration} right). It is obtained by sending the base point to infinity in the horizontal direction. The frame pseudolines are $k$ horizontal lines, and the $n-k$ working pseudolines are vertical wires. The orders of the working pseudolines on a horizontal line sweeping the wiring diagram from top to bottom form the so-called \defn{allowable sequence} of the working arrangement, as defined in~\cite{GoodmanPollack}.


\subsection{Description of the algorithm}
\label{subsec:description}

Our algorithm can enumerate all topological \conf{n}{k}s up to either topological or combinatorial equivalence. In order to maintain a reasonable computation space and time, the main idea is to focus on the relative positions of the points of the configurations and to ignore at first the relative positions of the other crossings among the pseudolines. In other words, to work modulo mutation equivalence as defined in Section~\ref{subsec:equivrelations}.

More precisely, we first enumerate at least one representative of each mutation equivalence class of topological \conf{n}{k}s. From these representatives, we can derive:
\begin{enumerate}
\item all topological \conf{n}{k}s up to topological equivalence: we explore each connected component of the mutation graph with our representatives as starting nodes. 
\item all combinatorial \conf{n}{k}s that are topologically realizable: we reduce the result modulo combinatorial equivalence.
\end{enumerate}
Since our motivation is to study geometric \conf{n}{k}s, we are only interested by point (2). We discuss a relatively efficient approach to test combinatorial equivalence in Section~\ref{subsec:reduction}. In this section, we give details on the different steps in our algorithm.

\paragraph{Sweeping process}
Our algorithm sweeps the projective plane to construct a topological \conf{n}{k}. The \defn{sweep line} sweeps the configuration from the base line on the top of the disk to the base line on the bottom of the disk. Inside each frame region, it always passes through the base point and always completes the configuration into an arrangement of $n+1$ pseudolines. When it switches from one frame region to the next one, it coincides with the separating frame pseudoline. Along the way, it sweeps completely all the working pseudolines. Except those located on the frame pseudolines, we assume that the crossings of the configuration are reached one after the other by the sweep line. After the sweep line swept a crossing, we remember the order of its intersections with the working pseudolines. In other words, the sweeping process provides us with the allowable sequence of the working pseudolines of our configuration.

Since admissible mutations are irrelevant for us, we only focus on the moments when our sweep line sweeps a \cross{k}. Thus, two different events can occur:
\begin{itemize}
\item when the sweep line sweeps a working \cross{k}, and
\item when the sweep line sweeps a frame pseudoline.
\end{itemize}
In the later case, we sweep simultaneously $k-1$ frame \cross{k}s (each involving the frame pseudoline and $k-1$ working pseudolines), and $n-1-k(k-1)$ frame \cross{2}s (each involving the frame pseudoline and a working pseudoline). Between two such events, the sweep line may sweep working \cross{2}s which are only taken into account when we reach a new event. Let us repeat again that the precise positions of these working \cross{2}s is irrelevant in our enumeration.

To obtain all possible solutions, we maintain a stack with all subconfigurations which have been constructed so far, remembering for each one:
\begin{enumerate}[(i)]
\item the order of the working pseudolines on the current sweep line, 
\item the number of frame and working \cross{k}s and \cross{2}s which have already been swept on each working pseudoline,
\item the length of the segment currently swept by the sweep line, and
\item the history of the sweeps performed to reach this subconfiguration.
\end{enumerate}
At each step, we remove the first subconfiguration from the stack, and insert all admissible subconfigurations which can arise after sweeping a new working \cross{k} or a new frame pseudoline. We finally accept a configuration once we have swept $k$ frame pseudolines and $n - k(k-1) - 1$ working \cross{k}s.

Any subconfiguration considered during the algorithm is a potential \conf{n}{k}. Throughout the process, we make sure that any pair of working pseudolines cross at most once, that the number of frame pseudolines (resp.~of working \cross{k}s) already swept never exceeds $k$ (resp.~$n-1-k(k-1)$), and that the total number of working \cross{2}s never exceeds $(n-2k)(n-1-k(k-1))/2$. Furthermore, on each pseudoline, the number of frame and working \cross{k}s (resp.~\cross{2}s) already swept never exceeds $k$ (resp.~${n-1-k(k-1)}$), the number of working $2$- and \cross{k}s already swept never exceeds ${n-1-k(k-1)}$, and the segment currently swept is not longer than the leftmost segment of the base~line.

We now detail individually each step of the algorithm.

\paragraph{Initialization}
We initialize our algorithm sweeping the base line. We only have to choose the distribution of the lengths of the segments on the base line. The possibilities are given by the list $\partitions$ of maximal representatives of $k$-partitions of $n-1-k(k-1)$.

\paragraph{Sweep a working \cross{k}}
If we decide to sweep a working \cross{k}, we have to choose the $k$ working pseudolines which intersect at this \cross{k}, and the direction of the other working pseudolines.

Since we are allowed to perform any admissible mutation, we can assume that all the pseudolines located to the left of the leftmost pseudoline of the working \cross{k}, and all those located to the right of the rightmost pseudoline of the working \cross{k} do not move. 

\begin{figure}
  \centerline{\includegraphics[scale=.7]{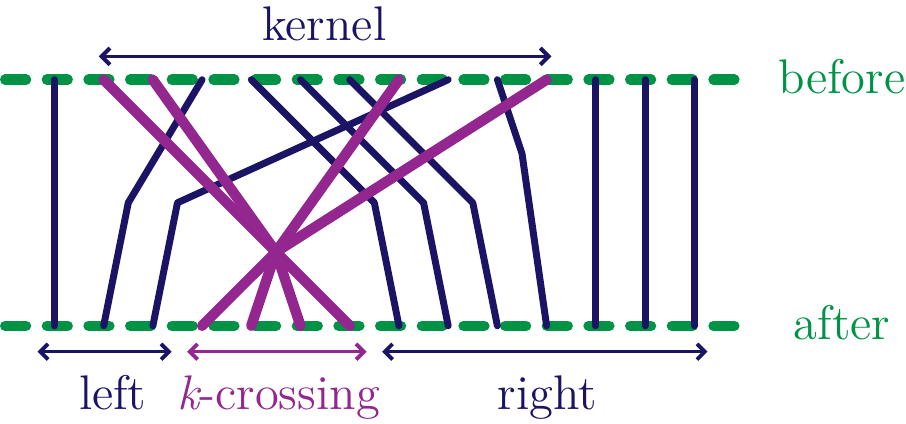}}
  \caption{Sweeping a working \cross{k}.}
  \label{fig:workingkCrossing}
\end{figure}

We say that the pseudolines located between the leftmost and the rightmost pseudolines of the working \cross{k} form the \defn{kernel} of the working \cross{k}. We have to choose the positions of the pseudolines of the kernel after the flip: each pseudoline of the kernel either belongs to the working \cross{k}, or goes to its left, or goes to its right (see Figure~\ref{fig:workingkCrossing}).

A choice of directions for the kernel is admissible provided that
\begin{enumerate}[(i)]
\item each pseudoline involved in the $k$-crossing can still accept a working $k$-crossing;
\item each pseudoline of the kernel can still accept as many working \cross{2}s as implied by the choice of directions for the kernel;
\item no segment becomes longer than the leftmost segment of the base line; and
\item any two pseudolines which are forced to cross by the choice of directions for the kernel did not cross earlier (\ie they still form an inversion on the sweep line before we sweep the working \cross{k}).
\end{enumerate}

\paragraph{Sweep a frame pseudoline}
If we decide to sweep a frame pseudoline, we have to choose the $(k-1)^2$ working pseudolines involved in one of the $k-1$ frame \cross{k}s, and the direction of the other working pseudolines.

As before, we can assume that a pseudoline does not move if it is located to the left of the leftmost pseudoline involved in one of the $k-1$ frame \cross{k}s, or to the right of the rightmost pseudoline involved in one of the $k-1$ frame \cross{k}s. Otherwise, we can perform admissible mutations to ensure this situation.

\begin{figure}
  \centerline{\includegraphics[scale=.7]{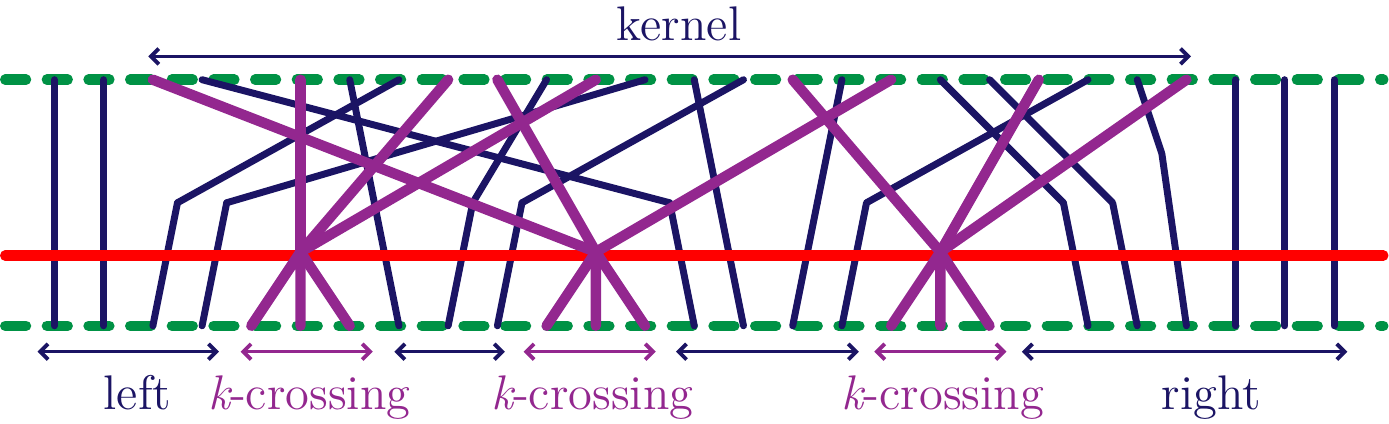}}
  \caption{Sweeping a frame pseudoline (right).}
  \label{fig:frameSweep}
\end{figure}

The other pseudolines form again the \defn{kernel} of the frame sweep, and we have to choose their positions after the flip. Each pseudoline of the kernel either belongs to one of the $k-1$ frame \cross{k}s, or can choose among $k$ possible directions: before the first frame \cross{k}, or between two consecutive frame \cross{k}s, or after the last frame \cross{k} (see Figure~\ref{fig:frameSweep}).

As before, a choice of directions for the kernel is admissible if 
\begin{enumerate}[(i)]
\item each pseudoline involved (resp. not involved) in one of the $k-1$ frame \cross{k}s can still accept a frame \cross{k} (resp. a frame \cross{2}); \item each pseudoline of the kernel can still accept as many working \cross{2}s as implied by the choice of directions for the kernel;
\item no segment becomes longer than the leftmost segment of the base line; and
\item any two pseudolines which are forced to cross by the choice of directions for the kernel did not cross earlier (\ie they still form an inversion on the sweep line before we sweep the frame pseudoline).
\end{enumerate}

\paragraph{Sweep the last frame region}
Our sweeping process finishes once we have swept $n-1-k(k-1)$ working \cross{k}s and $k$ frame pseudolines.
Each resulting subconfiguration should still be completed into a topological \conf{n}{k} with some necessary remaining \cross{2}s.
More precisely, we need to add on each working pseudoline as many working \cross{2}s as its number of inversions in the permutation given by the working pseudolines on the final sweep line, without creating segments that are too long.

After this last selection, all the constructed configurations are finally guaranteed to be valid topological \conf{n}{k}s. To make sure that we indeed obtain the representation presented in Section~\ref{subsec:representation}, we remove each configuration~$(P,L)$ in which the maximal representative of the segment length distribution of a pseudoline of $L$ appears in the list $\partitions$ before the segment length distribution of its base line.


\subsection{Testing combinatorial equivalence}
\label{subsec:reduction}

In Section~\ref{subsec:equivrelations}, we have seen three equivalence relations between topological \conf{n}{k}s: combinatorial, mutation and topological equivalence. As explained in Section~\ref{subsec:description}, our algorithm outputs at least one representative per mutation equivalence class of topological \conf{n}{k}s. However, we can obtain more than one representative per~class, and two topological \conf{n}{k}s which are not mutation equivalent can still be combinatorially equivalent. We thus need to reduce the output of our algorithm. 

\enlargethispage{.1cm}
Note that the topological equivalence between two \conf{n}{k}s~$(P_1,L_1)$ and~$(P_2,L_2)$ can be tested in $\Theta(n^3)$ time. Indeed, since the topological configurations are embedded on the projective plane, the matchings between~$P_1$ and~$P_2$ and between~$L_1$ and~$L_2$ induced by an homeomorphism mapping~$(P_1,L_1)$ to~$(P_2,L_2)$ are determined by the images of any two distinguished pseudolines~$\ell,\ell'$ of~$L_1$. Therefore, for each of the $\Theta(n^2)$ possible choices for the images of~$\ell,\ell'$, we can test in linear time whether this choice yields or not an homeomorphism between~$(P_1,L_1)$ and~$(P_2,L_2)$.
Both combinatorial and mutation equivalences are however harder to decide computationally. We focus here on methods and heuristics to quickly test combinatorial~equivalence.

In order to limit unnecessary computation, we make use of \defn{combinatorial invariants} associated to configurations. If two configurations have distinct invariants, they cannot be combinatorially equivalent. Reciprocally, if they share the same invariant, it provides us with information on the possible combinatorial isomorphisms between these two configurations. The invariants we have chosen are the \defn{clique} and \defn{coclique distributions}. We furthermore need a \defn{multiscale invariant} technique, based on the notion of \defn{derivation} of a combinatorial invariant. We introduce these notions and methods in the next paragraphs.

\paragraph{Cliques and cocliques}
Let $(P,L)$ be a combinatorial configuration.
For $j\ge 3$, define a \defn{$j$-clique} of~$(P,L)$ to be any set of $j$ points of~$P$ which are pairwise related by lines of~$L$. For any point $p$ of $P$, let $\gamma_j(p)$ be the number of $j$-cliques containing $p$, and let $\gamma(p) := (\gamma_j(p))_{j \ge 3}$. The \defn{clique distribution} of~$(P,L)$ is the multiset $\gamma(P) := \multiset{\gamma(p)}{p \in P}$.

Similarly, a \defn{$j$-coclique} of~$(P,L)$ is a set of $j$ lines of~$L$ which are pairwise intersecting at points of~$P$. For any line $\ell$ of $L$, let $\delta_j(\ell)$ be the number of $j$-cocliques containing $\ell$, and let $\delta(\ell) := (\delta_j(\ell))_{j \ge 3}$. The \defn{coclique distribution} of~$(P,L)$ is the multiset $\delta(L) := \multiset{\delta(\ell)}{\ell \in L}$. In other words, the coclique distribution of $(P,L)$ is the clique distribution of its dual configuration~$(L,P)$.

The pair $(\gamma(P), \delta(L))$ of clique and coclique distributions of the configuration~$(P,L)$ is a natural and powerful combinatorial invariant of~$(P,L)$.

\paragraph{Derivation of combinatorial invariants}
Let~$(P,L)$ be a \conf{n}{k}. Assume that $\gamma:P\to X$ and $\delta:L\to Y$ are two functions from the point set and the line set of~$(P,L)$ respectively to arbitrary sets $X$ and $Y$, such that the multisets $\gamma(P) := \multiset{\gamma(p)}{p \in P} \subset X$ and $\delta(L) := \multiset{\delta(\ell)}{\ell \in L} \subset Y$ are combinatorial invariants of~$(P,L)$. The clique and coclique distributions are typical examples of such functions~$\gamma$ and~$\delta$. Observe that we again abuse notation: the functions~$\gamma$ and~$\delta$ usually depend on the configuration~$(P,L)$, but we consider that this dependence is clear from the context. Note however that the target sets~$X$ and~$Y$ of $\gamma$ and $\delta$ do not depend upon~$(P,L)$.

While reducing a set of configurations up to combinatorial equivalence, such a pair of combinatorial invariants~$(\gamma(P),\delta(L))$ can be used in two different ways:
\begin{enumerate}[(i)]
\item either to separate classes of combinatorial isomorphism: two configurations with different invariants cannot be combinatorially equivalent;
\item or to guess combinatorial isomorphisms: an isomorphism between two configurations should respect the invariants~$\gamma$ and~$\delta$.
\end{enumerate}
It often happens however that the pair of combinatorial invariants~$(\gamma(P),\delta(L))$ is not precise enough neither to distinguish two configurations, nor to guess a combinatorial isomorphism between them. It occurs when many points (resp. many lines) of a configuration~$(P,L)$ get the same image under~$\gamma$ (resp. under $\delta$). Two fundamentally different cases can lead to this situation. On the one hand, the configuration~$(P,L)$ can have a large automorphism group. In this case, points (resp. lines) in a common orbit under the automorphism group cannot be distinguished combinatorially, and thus no invariant can speed up the isomorphism test. On the other hand, it could also be that the combinatorial invariant~$(\gamma(P),\delta(L))$ is not precise enough to distinguish the neighborhood properties of the points with the same image under $\gamma$ (resp. the lines with the same image under $\delta$). In the later case, we can construct a new pair of combinatorial invariants which refines~$(\gamma(P),\delta(L))$, taking into account the neighborhoods of points and lines in the configuration. We call these invariants the \defn{derivatives} of $\gamma$ and $\delta$ and denote them $\gamma'$ and $\delta'$.

The \defn{derivative} of the invariant $\gamma : P \to X$ is the function $\gamma':L\to X^k$ which associates to a line $\ell$ of~$L$ the multiset $\gamma'(\ell) := \multiset{\gamma(p)}{p\in P, p\in \ell}$. Intuitively, the image $\gamma'(\ell)$ of a line $\ell$ contains all the combinatorial information carried by~$\gamma$ concerning the points of~$P$ contained in $\ell$. Similarly, the \defn{derivative} of the invariant $\delta : L \to Y$ is the function $\delta':P\to Y^k$ which associates to a point $p$ of $P$ the multiset ${\delta'(p) := \multiset{\delta(\ell)}{\ell \in L, p\in \ell}}$. The pair $(\delta'(P), \gamma'(L))$ is a pair of combinatorial invariants as defined previously, and it refines the previous pair~$(\gamma(P),\delta(L))$. 

If this new invariant is still not precise enough, we can consider higher order derivatives $\gamma^{(u)} := (\gamma^{(u-1)})'$ and $\delta^{(u)} := (\delta^{(u-1)})'$ of the initial invariants. We obtain this way a family of refinements of $(\gamma(P),\delta(L))$. Of course, these invariants ultimately carry the same combinatorial information. We use this family in the following multiscale technique.

\paragraph{Multiscale invariants}
The main idea of our reduction process is to use derivative invariants in a multiscale process. Consider a set~$\cC$ of configurations that we want to reduce up to combinatorial equivalence. Assume that $\gamma:P\to X$ and $\delta:L\to Y$ are two functions defining a pair of combinatorial invariants~$(\gamma(P),\delta(L))$ of a configuration~$(P,L)$. We separate the configurations of~$\cC$ into classes with distinct invariants, which we can consider independently. We now compute the derivative invariants $(\delta'(P), \gamma'(L))$ for each configuration~$(P,L)$. For a given class, we then have three possible situations:
\begin{enumerate}
\item If the derivative invariants $(\delta'(P), \gamma'(L))$ are not the same for all configurations $(P,L)$ of the class, we split the class into refined subclasses and reiterate the refinement (computing one more derivative).
\item If the derivative invariants $(\delta'(P), \gamma'(L))$ are the same for all configurations $(P,L)$ of the class but determine more information on the possible isomorphisms between configurations of the class than the original invariants~$(\gamma(P), \delta(L))$, then we reiterate the refinement.
\item Otherwise, the derivative invariants $(\delta'(P), \gamma'(L))$, as well as any further derivative, provide the same combinatorial information as the original invariants $(\gamma(P), \delta(L))$. Thus, we stop the refinement process and start a brute-force search for possible isomorphisms between the remaining configurations in the class. The efficiency of this brute-force search depends on the quality of the combinatorial information provided by the invariants~$(\gamma(P), \delta(L))$.
\end{enumerate}
This process can be seen as a multiscale process: typically, some invariants provide sufficiently information to deal with certain classes of~$\cC$, while other classes require far more precision (obtained by derivatives) to be reduced.

Using this multiscale technique, starting from the clique and coclique distributions of configurations, we managed to reduce the $69\,991$ topological \conf{19}{4}s produced by our sweeping algorithm into $4\,028$ classes of combinatorial equivalence in about one hour\footnote{Computation times on a 2.4 GHz Intel Core 2 Duo processor with 4Go of RAM.}.


\subsection{Results}
\label{subsec:resultsTopo}

\begin{figure}[h]
  \centerline{\includegraphics[scale=.55]{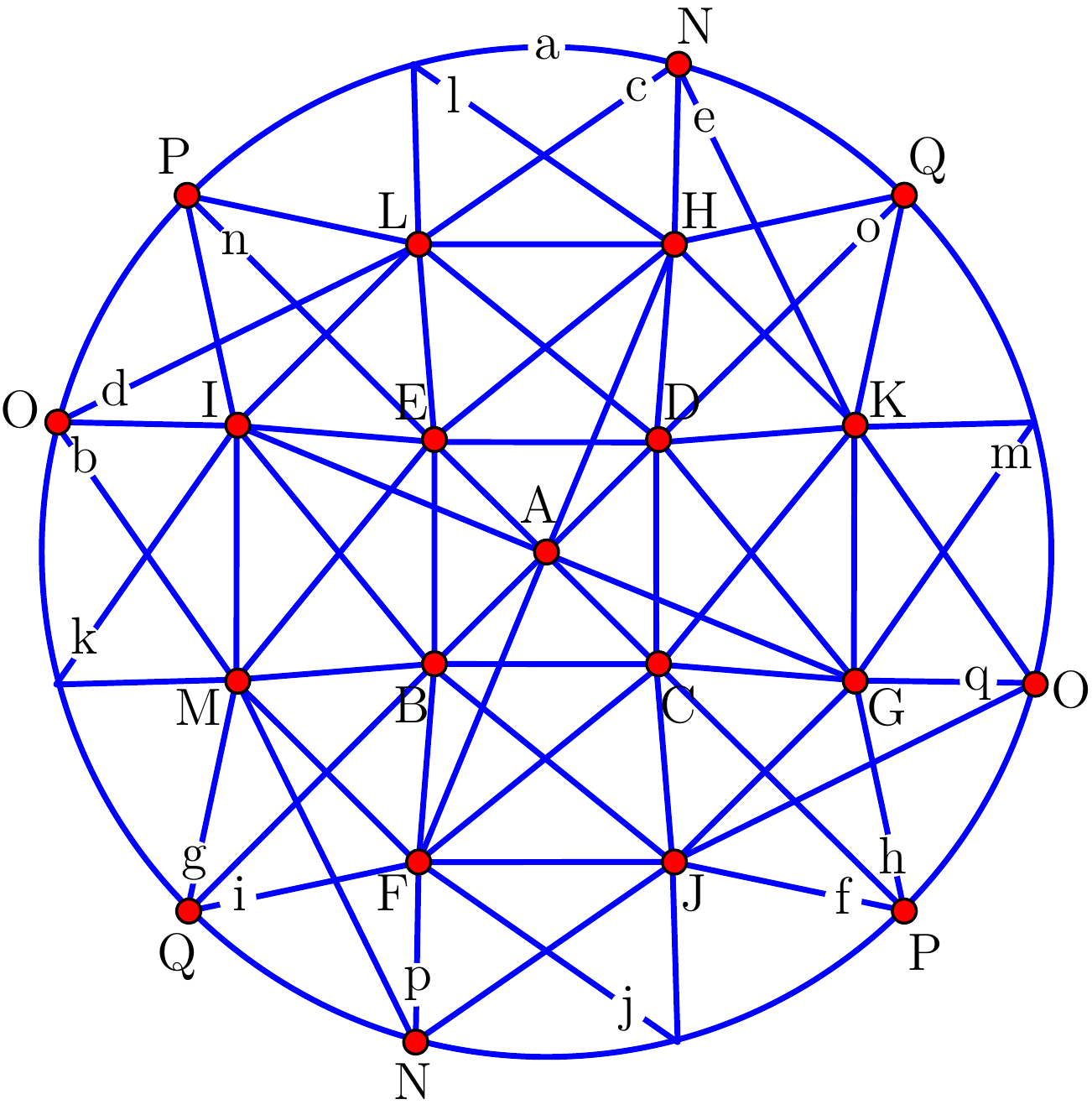}}

  \medskip
  \centerline{
	\begin{tabular}{c|cccccccccccccccccc}
	lines 	& a & b & c & d & e & f & g & h & i & j & k & l & m & n & o & p & q \\
	\hline
			& N & F & I & H & G & B & E & D & C & B & E & D & C & A & A & A & A \\
 	points	& P & J & M & L & K & I & H & G & F & E & D & C & B & C & B & F & G \\
 	in lines	& O & O & N & O & N & P & Q & P & Q & L & K & J & M & P & Q & N & O \\
   			& Q & M & L & K & J & J & M & L & K & F & I & H & G & E & D & H & I
	\end{tabular}
  }
  \caption{The topological \conf{17}{4}~\cite{BokowskiGrunbaumSchewe}.}
  \label{fig:configuration_17_4}
\end{figure}

We present in this section the results of our algorithm. First, it enables us to check efficiently all former enumerations of topological \conf{n}{k}s. The \Java{} implementation developed by the second author finds all \conf{n}{k}s in less than a minute\textsuperscript{1} when $k=3$ and $n\le11$, or when $k=4$ and~${n\le 17}$. In particular, we checked that there is no topological \conf{n}{4} when $n\le 16$~\cite{BokowskiSchewe1}, and that there is a single topological \conf{17}{4} up to combinatorial isomorphism. This configuration is represented in Figure~\ref{fig:configuration_17_4}, and labeled in such a way that:
\begin{itemize}
\item the quarter-turn rotation which generates the symmetry group of the picture is the permutation (A)(B,C,D,E)(F,G,H,I)(J,K,L,M)(N,O)(P,Q); and
\item the permutation (A,a)(B,b)\,\dots\,(P,p)(Q,q) is a self-polarity of the topological configuration.
\end{itemize}

When $k=4$ and $n=18$, we reconstructed the $16$ combinatorial equivalence classes of topological \conf{18}{4}s obtained in~\cite{Schewe} with satisfiability solvers. See~\cite[Figure~6]{BokowskiSchewe2} for a description of these configurations. To obtain this result, our implementation needed about one hour\footnote[1]{Computation times on a 2.4 GHz Intel Core 2 Duo processor with 4Go of RAM.}, compared to several months of CPU-time required in~\cite{Schewe}. The two \conf{18}{4}s presented in Figure~\ref{fig:configurations_18_4}, which are combinatorially equivalent but not mutation equivalent, occurred while we were reducing the list of \conf{18}{4}s up to combinatorial equivalence, using as a first reduction a certain invariant of mutation equivalence defined in~\cite{BokowskiStrauszSantiago}. In the next section, we present two combinatorially distinct geometric \conf{18}{4}s obtained from the list of topological \conf{18}{4}s.

Finally, we want to report on preliminary results concerning the enumeration of topological \conf{19}{4}s, which initially motivated our work. In about $15$ days of computation time\textsuperscript{1}, we obtained the complete list of topological \conf{19}{4}s:

\begin{resultat}
There are precisely $4\,028$ topological \conf{19}{4}s up to combinatorial equivalence. Among them, $222$ are self-dual.
\end{resultat}

From this list, we can immediately extract examples of topological \conf{19}{4}s with non-trivial symmetry groups, closing along the way an open question of Branko Gr\"unbaum~\cite[p.~169, Question~5]{Grunbaum1}. The next step is naturally to study the possible geometric realizations of all these topological \conf{19}{4}s. This work in progress still requires an important computational effort and will be reported in a subsequent paper.


\section{Application to geometric \conf{18}{4}s}
\label{sec:geomconf}

As an application of the enumeration of topological configurations, we derive all isomorphism classes of geometric \conf{18}{4}s. To obtain it, we implemented in \Maple{} the construction sequence method of J\"urgen Bokowski and Lars Schewe~\cite{BokowskiSchewe2}. Among the $16$ topological \conf{18}{4}s (first generated by Lars Schewe~\cite{Schewe} and now confirmed by our \Java{} program), only $8$ are compatible with Pappus' and Desargues' Theorem. Starting from these remaining configurations, we run our \Maple{} code and obtain the following result:

\begin{resultat}
There are precisely two geometric \conf{18}{4}s up to combinatorial isomorphism.
\end{resultat}

The first geometric \conf{18}{4} was obtained in~\cite[Section~4]{BokowskiSchewe1}.

\begin{figure}[b]
  \centerline{\includegraphics[scale=.31]{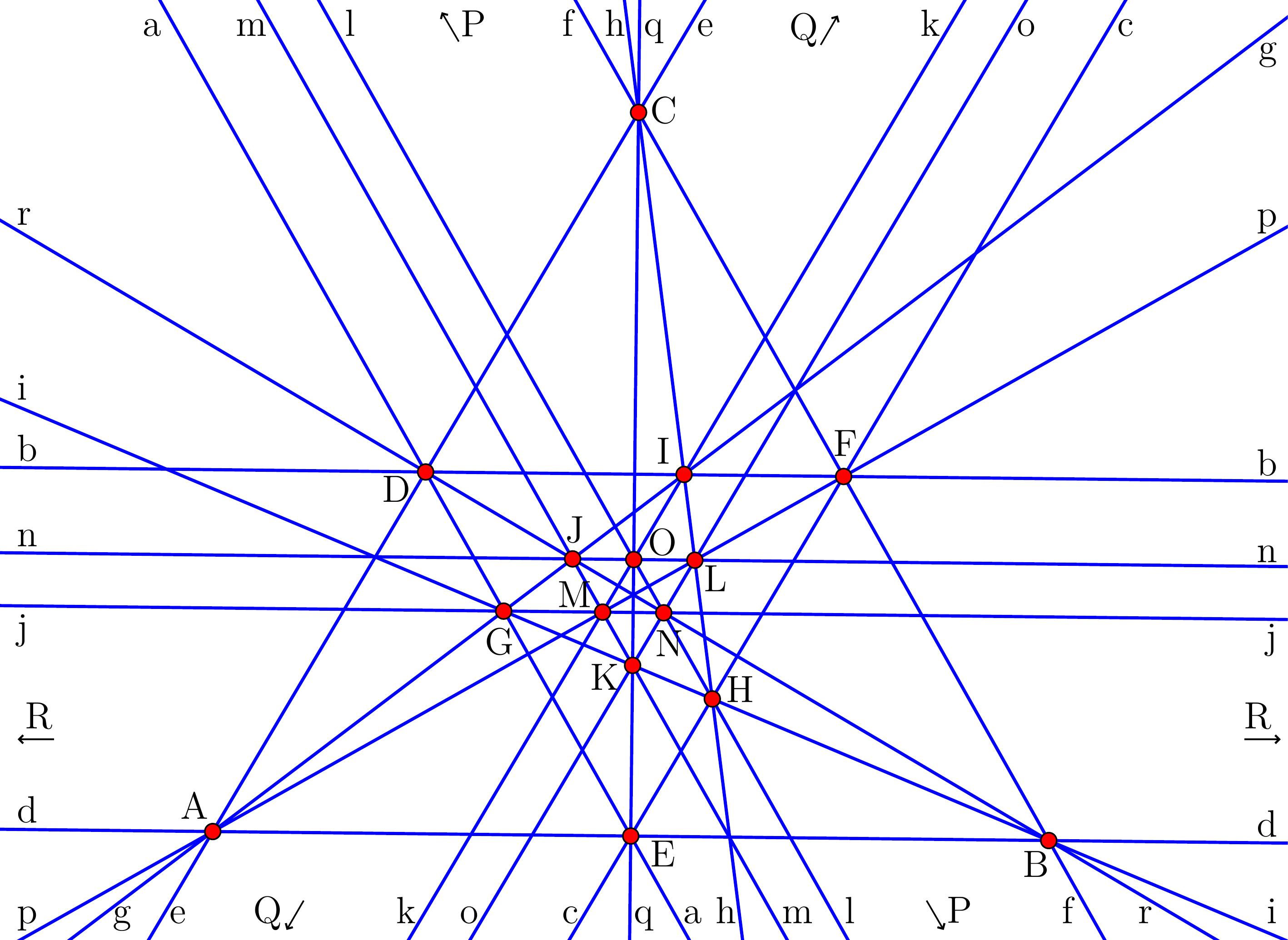}\qquad\includegraphics[scale=.54]{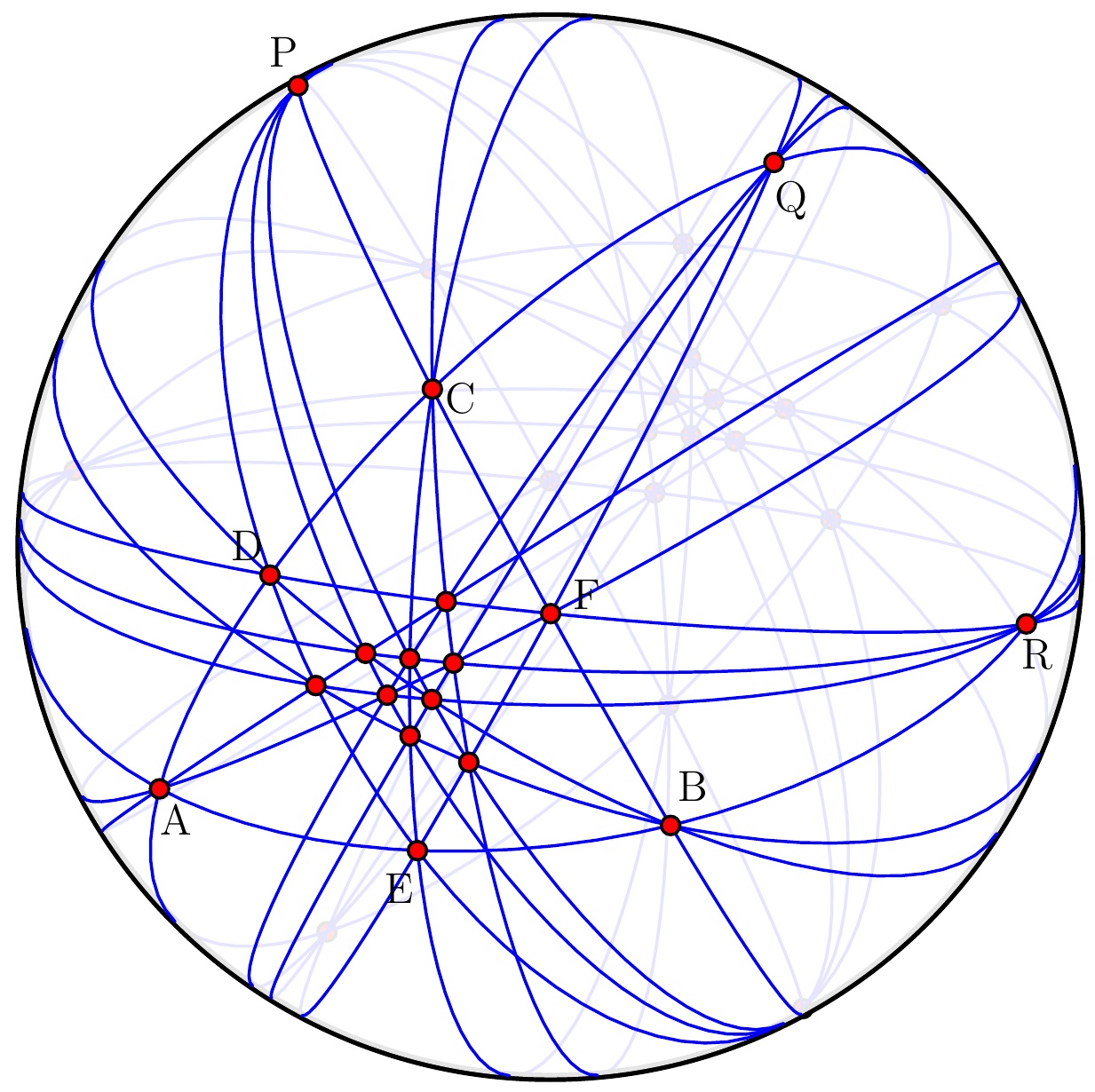}}
  
  \bigskip
	\begin{tabular}{c|cccccccccccccccccc}
	lines 	& a & b & c & d & e & f & g & h & i & j & k & l & m & n & o & p & q & r \\
	\hline
			& D & F & E & A & C & B & A & C & B & G & I & H & J & L & K & A & C & B \\
	points	& G & I & H & E & D & F & G & I & H & M & O & N & M & O & N & F & E & D \\
	in lines	& E & D & F & B & A & C & J & L & K & N & M & O & K & J & L & L & K & J \\
  			& P & R & Q & R & Q & P & I & H & G & R & Q & P & P & R & Q & M & O & N
	\end{tabular}

  \caption{Bokowski and Schewe's geometric \conf{18}{4}~\cite{BokowskiSchewe1}.}
  \label{fig:firstConfiguration184}
\end{figure}

\begin{figure}[h]
  \centerline{\includegraphics[scale=.31]{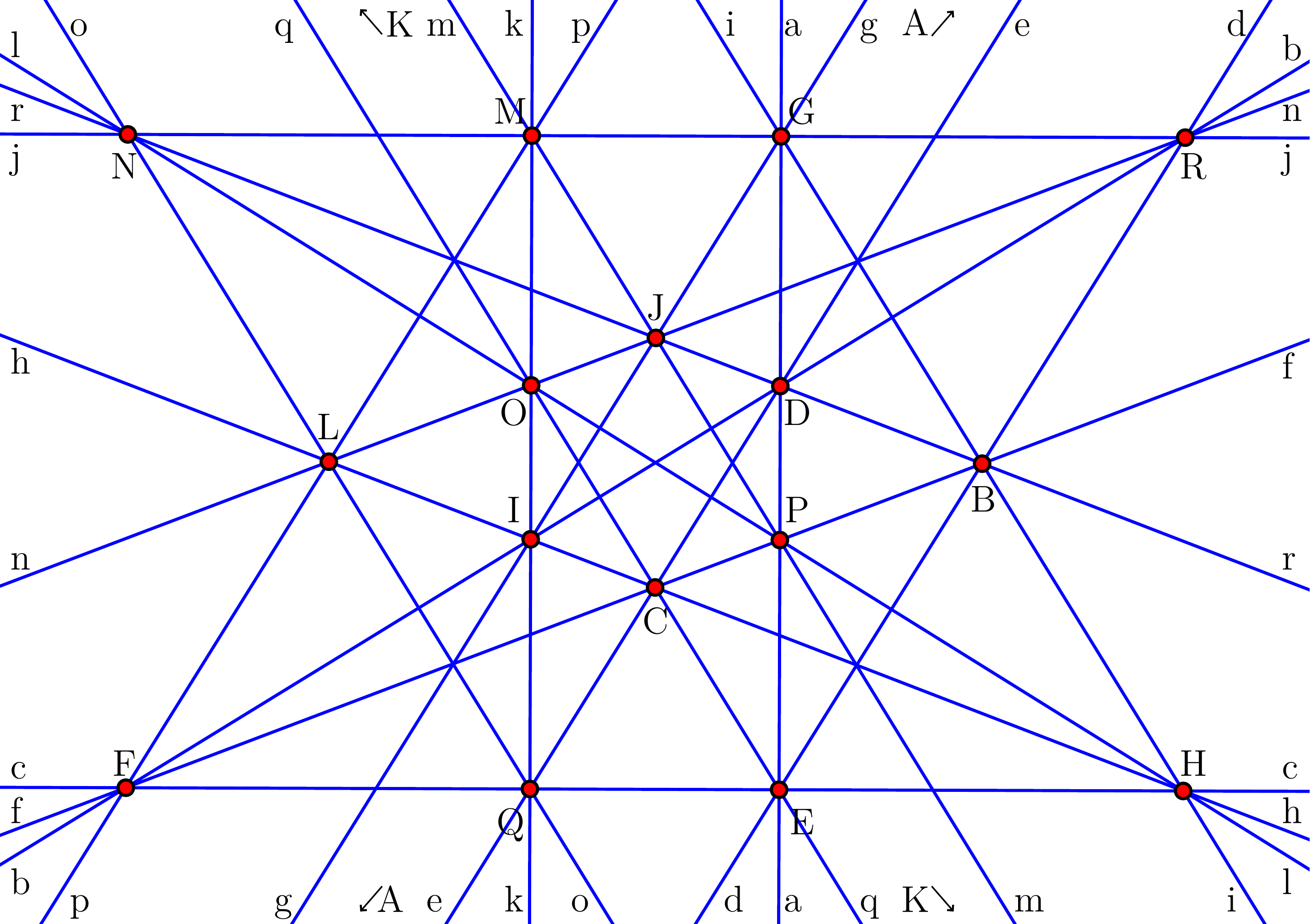}\qquad\includegraphics[scale=.54]{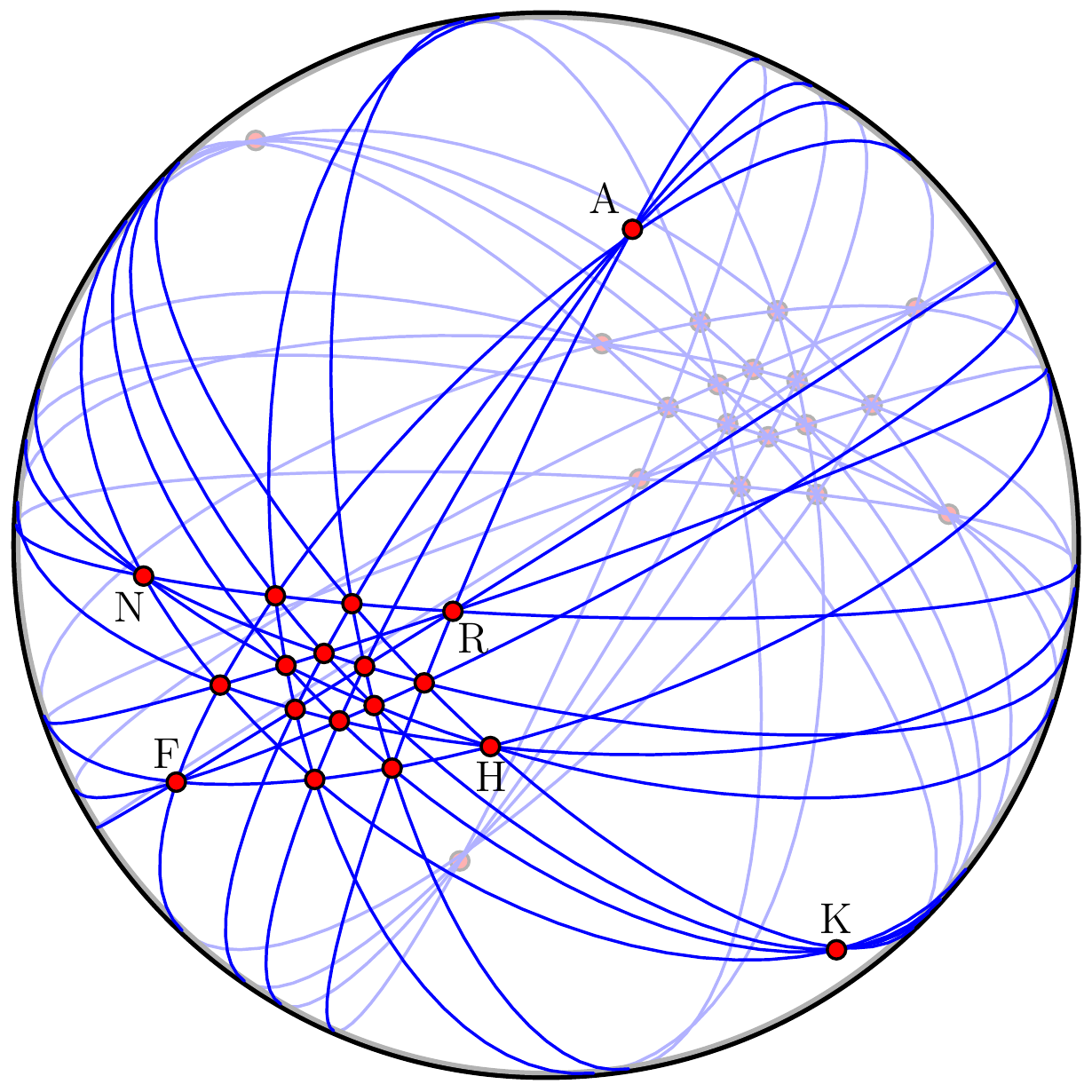}}
  \caption{Another geometric realization of Bokowski and Schewe's geometric \conf{18}{4}~\cite{BokowskiSchewe1} of Figure~\ref{fig:firstConfiguration184}.}
  \label{fig:firstConfiguration184new}
\end{figure}

In Figure~\ref{fig:firstConfiguration184}, we have labeled its points A,\,\dots,\,R and lines a,\,\dots,\,r in such a way that the permutation (A,a)(B,b)\,\dots\,(Q,q)(R,r) is a self-duality of the configuration. 
The automorphism group of the combinatorial configuration is generated by the permutations:
\begin{center}
(A,B,C)(D,E,F)(G,H,I)(J,K,L)(M,N,O)(P,Q,R) \\ 
(A)(K)(B,C,L,J)(D,F,I,R)(E,Q,M,G)(H,O,N,P) \\ 
(A)(K)(B,L)(C,J)(D,I)(E,M)(F,R)(G,Q)(H,N)(O,P) \\ 
\end{center}
and is isomorphic to the symmetric group on $4$ elements. Together with the self-duality (A,a)(B,b)\,\dots\,(Q,q)(R,r), the automorphism group of the Levi graph of the configuration is thus isomorphic to $\fS_4\times\Z_2$. Observe that only the first permutation (A,B,C)\,\dots\,(P,Q,R) and the duality (A,a)(B,b)\,\dots\,(Q,q)(R,r) are geometrically visible, while the other generators of the automorphism group of the combinatorial configuration are not isometries of the geometric configuration of Figure~\ref{fig:firstConfiguration184}.
In Figure~\ref{fig:firstConfiguration184new}, we have performed a projective transformation of the configuration of Figure~\ref{fig:firstConfiguration184} (sending the four 3-valent points in Figure~\ref{fig:firstConfiguration184} to a square).
The last generator (A)(K)(B,L)\,\dots\,(O,P) then becomes a central symmetry in the new geometric \conf{18}{4}. 

The realization space of this configuration consists of two points, both expressed with coordinates in $\Q\left[1+\sqrt{5}\right]$.

The second geometric \conf{18}{4} is a result of our \Maple{} code and appears for the first time in this paper.

\begin{figure}
  \centerline{\includegraphics[scale=.31]{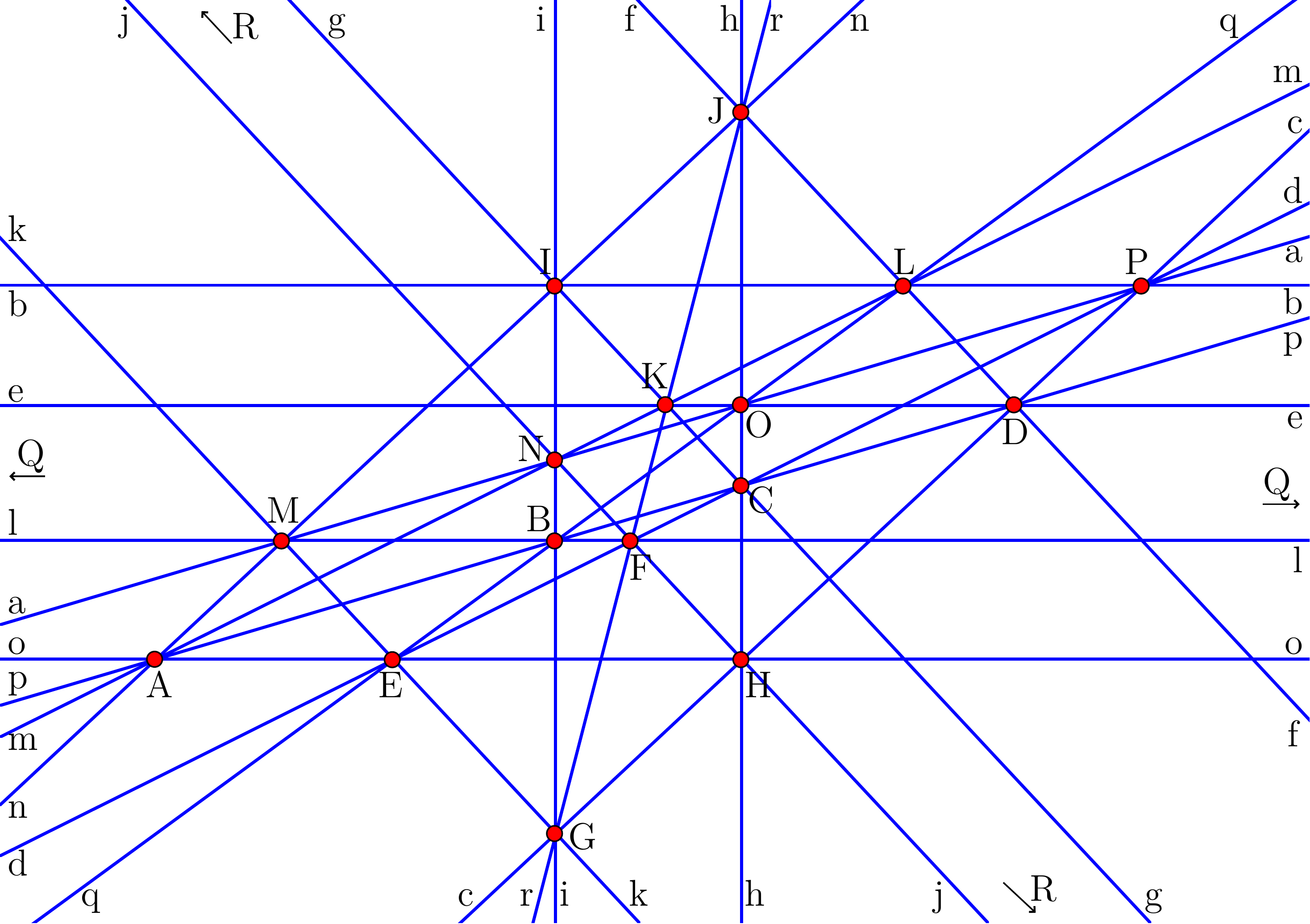}\qquad\includegraphics[scale=.54]{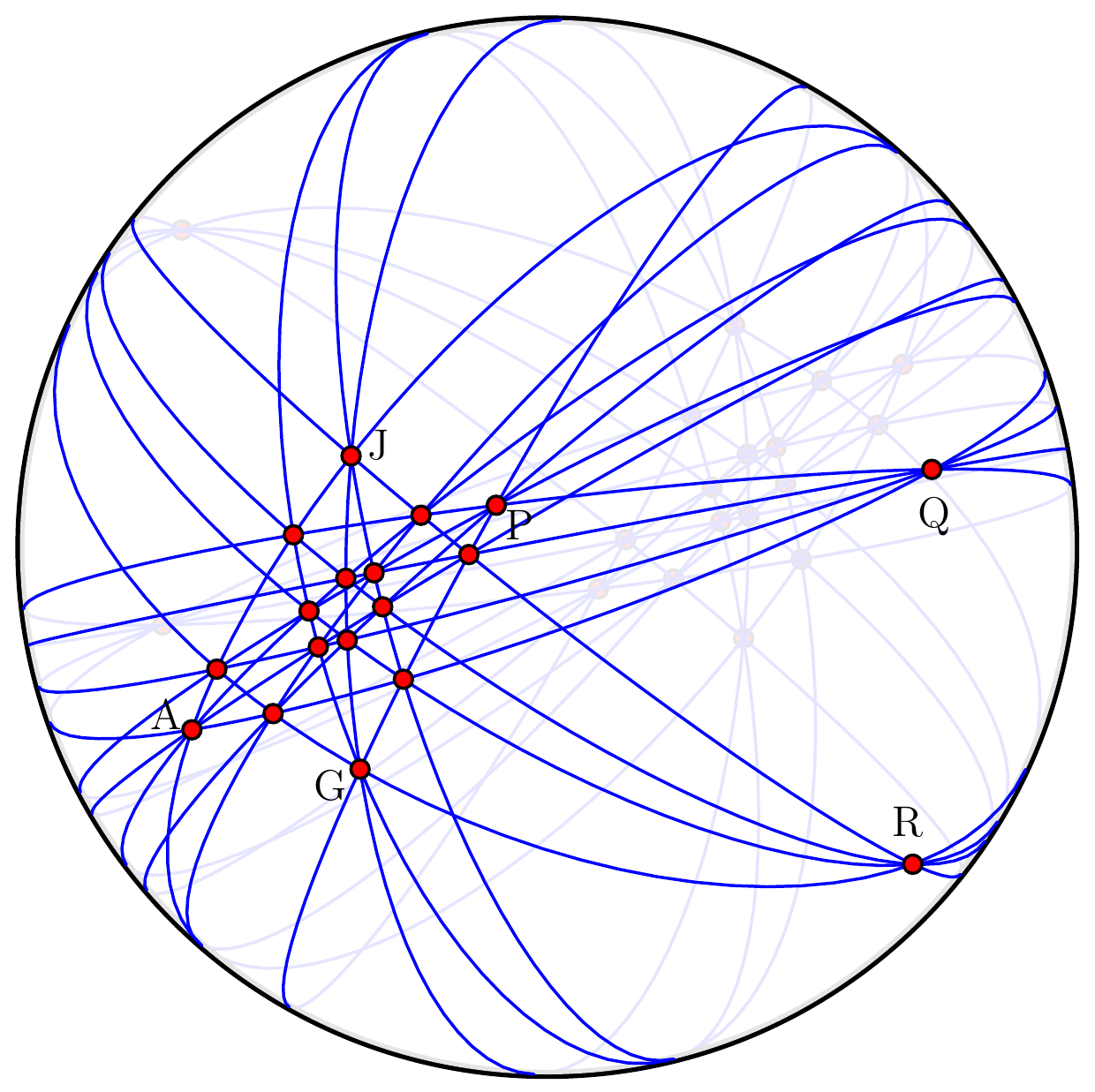}}

  \bigskip
  \centerline{
	\begin{tabular}{c|cccccccccccccccccc}
	lines 		& a & b & c & d & e & f & g & h & i & j & k & l & m & n & o & p & q & r \\
	\hline
  			& M & I & D & C & D & D & C & C & N & N & M & M & N & M & H & D & B & F \\
	points 	& N & L & H & F & O & L & K & H & I & F & E & B & K & I & E & C & E & G \\
	in lines	& O & P & G & E & K & J & I & J & G & H & G & F & L & J & A & B & L & J \\
  			& P & Q & P & P & Q & R & R & O & B & R & R & Q & A & A & Q & A & O & K
	\end{tabular}
  }
  \caption{The new geometric \conf{18}{4}.}
  \label{fig:secondConfiguration184}
\end{figure}

\enlargethispage{.1cm}
In Figure~\ref{fig:secondConfiguration184}, we have labeled its points A,\,\dots,\,R and lines a,\,\dots,\,r in such a way that the permutation (A,a)(B,b)\,\dots\,(Q,q)(R,r) is a self-polarity of the configuration.

The automorphism group of the combinatorial configuration is generated by the permutation (Q)(R)(A,P)(B,O)(C,N)(D,M)(E,L)(F,K)(G,J)(H,I). Together with the self-polarity (A,a)(B,b)\,\dots\,(Q,q)(R,r), the automorphism group of the Levi graph of the configuration is thus isomorphic to $\Z_2\times\Z_2$. This group is completely realized in the geometric representation of Figure~\ref{fig:secondConfiguration184}.

The realization space of this configuration consists of two points, both expressed with coordinates in $\Q\left[\sqrt[3]{108+12\sqrt{93}}\right]$

\begin{figure}[b]
  \centerline{\includegraphics[height=5cm]{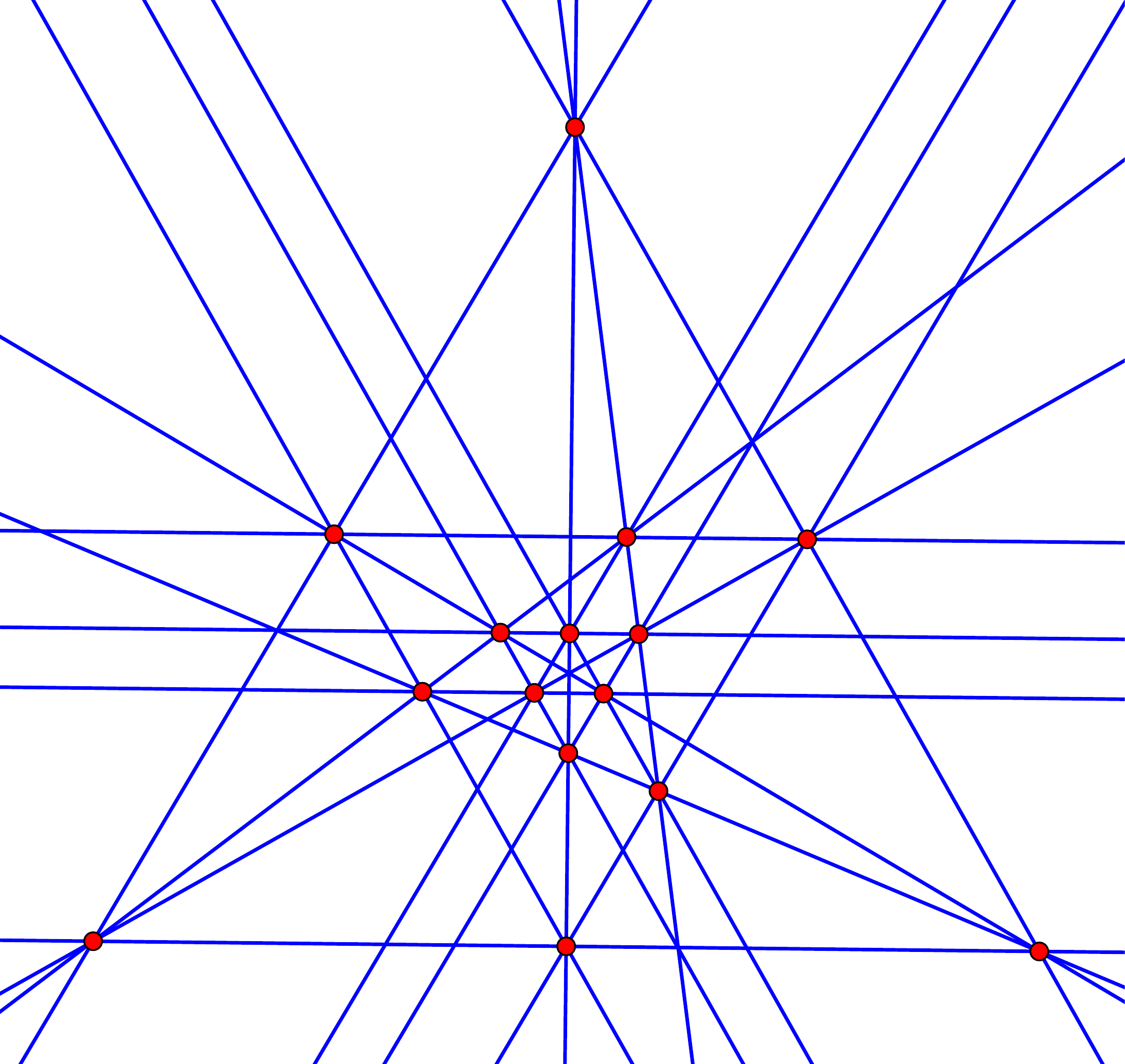}\quad\includegraphics[height=5cm]{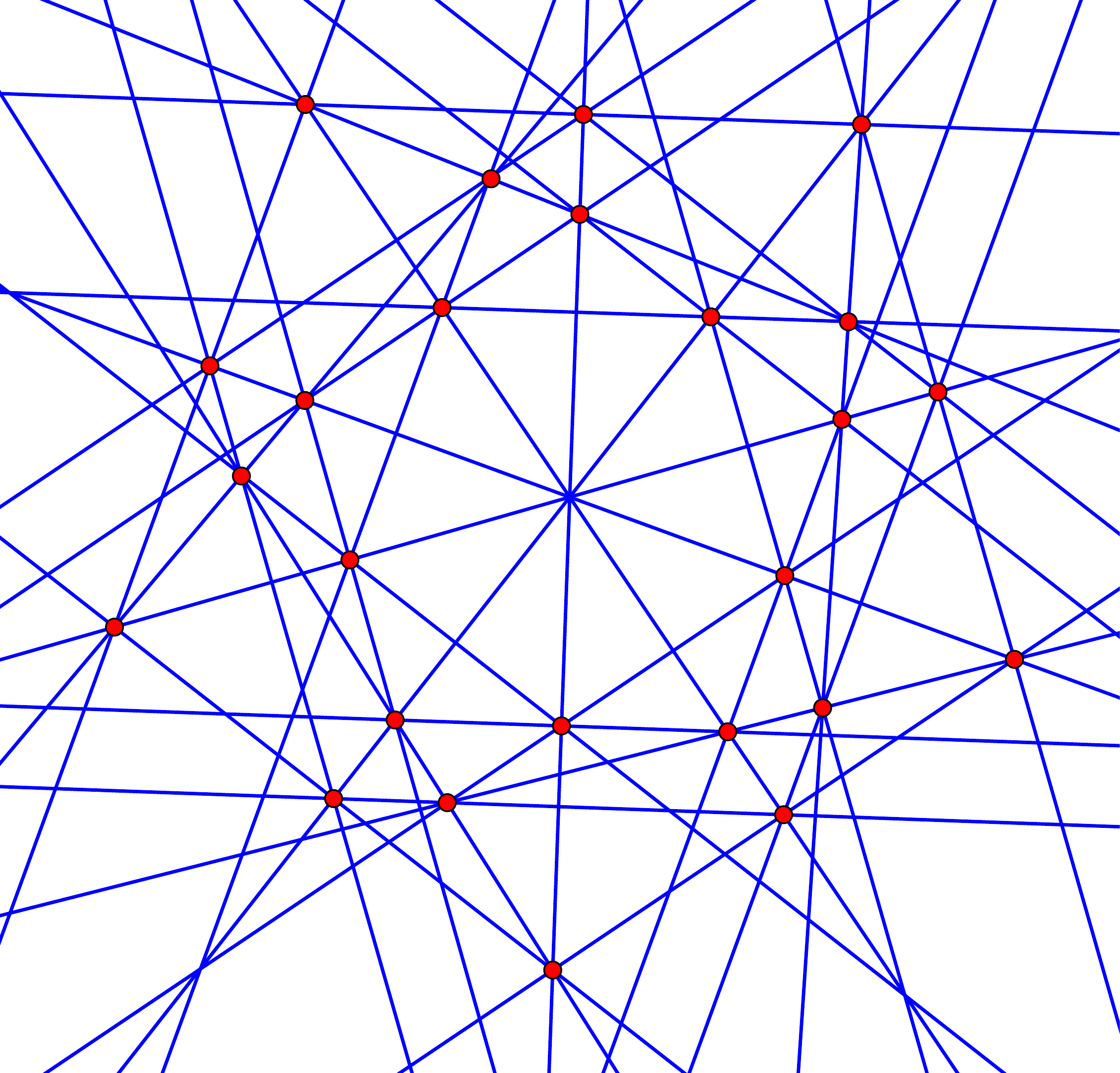}\quad\includegraphics[height=5cm]{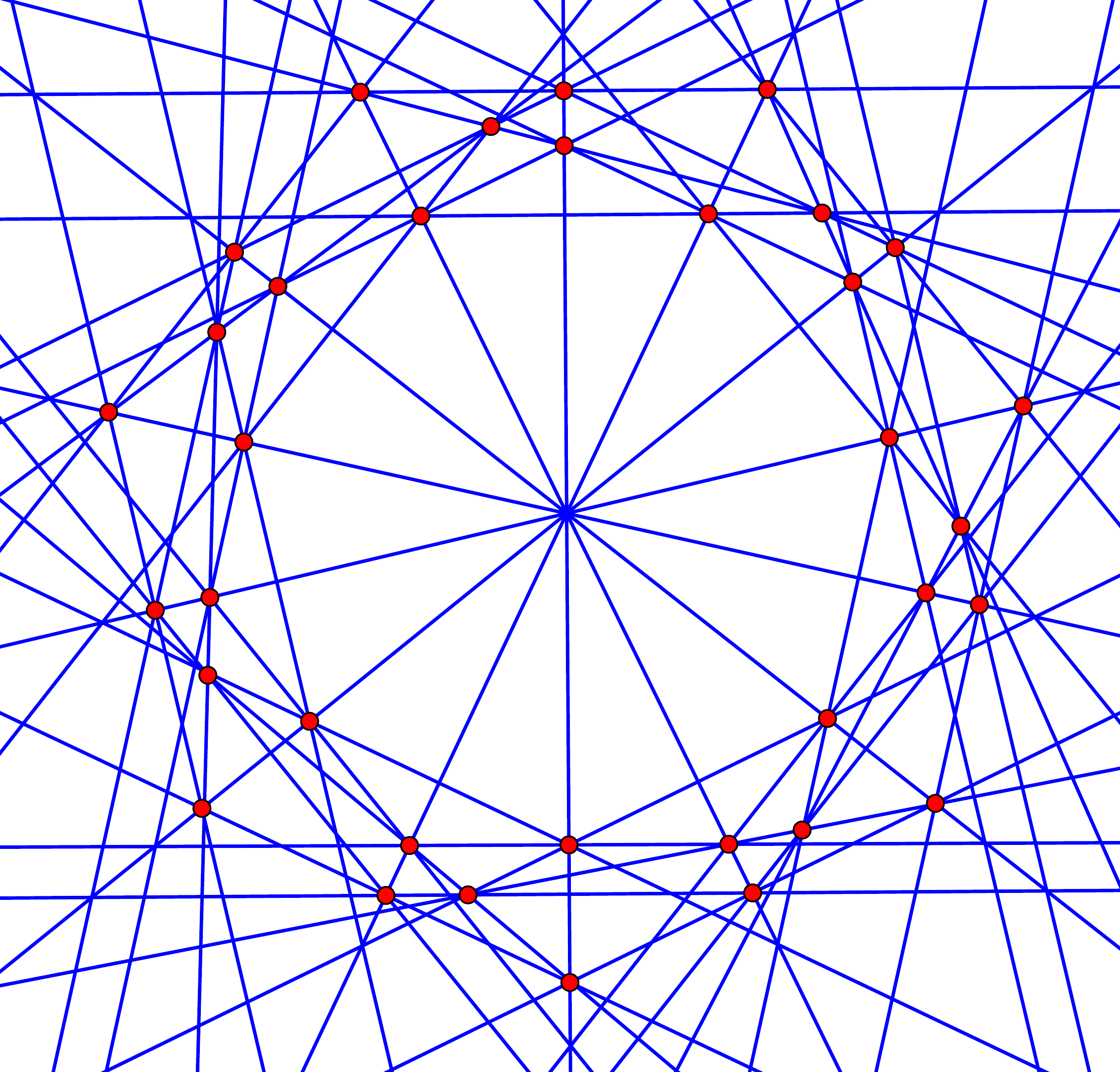}}
  \caption{The $(6m)$-family inspired by the geometric \conf{18}{4} of Figure~\ref{fig:firstConfiguration184}.}
  \label{fig:6mfamily}
\end{figure}

To conclude, we want to emphasize that the discovery of the first \conf{18}{4} of Figure~\ref{fig:firstConfiguration184} inspired Branko Gr\"unbaum to find a new family of \conf{6m}{4}s, for any $m \ge 3$ (see~\cite[Chapter~3, p.~171]{Grunbaum1} and Figure~\ref{fig:6mfamily}). This raises the following appealing open question:

\begin{probleme}
Generalize our second geometric \conf{18}{4} of Figure~\ref{fig:secondConfiguration184} to obtain another new infinite family of geometric \conf{n}{4}s.
\end{probleme}

For example, we have been able to derive from the second geometric \conf{18}{4} of Figure~\ref{fig:secondConfiguration184} a family of \conf{(18+17m)}{4}s. Unfortunately, the set~$18+17\N$ does not intersect the set~$\{19,22,23,26,37,43\}$ of values~$n$ for which no \conf{n}{4} is known.


\section*{Acknowledgements}

The first author thanks three colleagues from the Universidad Nacional Aut\'onoma de M\'exico, namely Ricardo Strausz Santiago, Rodolfo San Augustin Chi, and Octavio Paez Osuna, for many stimulating discussions about various different earlier versions of the presented algorithm during his one year sabbatical stay (2008/2009) in M\'exico City. We also thank Leah Berman from the University of Alaska Fairbanks for valuable discussions and comments about the subject. We are grateful to Branko Gr\"unbaum, Toma\v{z} Pisanski, and Gunnar Brinkmann for encouragements and helpful communications. As frequent users, we are indebted to the development team of the geometric software \Cinderella{}, in particular J\"urgen Richter-Gebert and Ulrich Kortenkamp. Finally, we thank two anonymous referees for their comments and suggestions on the presentation.


\bibliographystyle{alpha}
\bibliography{topologicalConfigurations.bib}

\end{document}